\documentclass[journal=NanoLetters,manuscript=article]{achemso}
\setkeys{acs}{keywords = true}
\usepackage{chemformula} 
\usepackage[T1]{fontenc} 
\usepackage{xcolor}
\usepackage{caption}
\usepackage{multirow}
\usepackage{color}
\usepackage{bm}
\usepackage{braket}
\usepackage{amsmath, amssymb, amsfonts}
\usepackage{calc}
\usepackage{siunitx}

\author{Xuanxi Cai$^{\infty}$}
\affiliation
{State Key Laboratory of Low Dimensional Quantum Physics and Department of Physics, Tsinghua University, Beijing 100084, P.R. China}
\author{Changhua Bao$^{\infty}$}
\affiliation
{State Key Laboratory of Low Dimensional Quantum Physics and Department of Physics, Tsinghua University, Beijing 100084, P.R. China}
\author{Benshu Fan$^{\infty}$}
\affiliation
{State Key Laboratory of Low Dimensional Quantum Physics and Department of Physics, Tsinghua University, Beijing 100084, P.R. China}
\alsoaffiliation
{Max Planck Institute for the Structure and Dynamics of Matter, Center for Free-Electron Laser Science, Hamburg 22761, Germany}
\author{Haoyuan Zhong}
\affiliation
{State Key Laboratory of Low Dimensional Quantum Physics and Department of Physics, Tsinghua University, Beijing 100084, P.R. China}
\author{Fei Wang}
\affiliation
{State Key Laboratory of Low Dimensional Quantum Physics and Department of Physics, Tsinghua University, Beijing 100084, P.R. China}
\author{Shaohua Zhou}
\affiliation
{State Key Laboratory of Low Dimensional Quantum Physics and Department of Physics, Tsinghua University, Beijing 100084, P.R. China}
\author{Tianyun Lin}
\affiliation
{State Key Laboratory of Low Dimensional Quantum Physics and Department of Physics, Tsinghua University, Beijing 100084, P.R. China}
\author{Hongyun Zhang}
\affiliation
{State Key Laboratory of Low Dimensional Quantum Physics and Department of Physics, Tsinghua University, Beijing 100084, P.R. China}
\author{Pu Yu}
\affiliation
{State Key Laboratory of Low Dimensional Quantum Physics and Department of Physics, Tsinghua University, Beijing 100084, P.R. China}
\alsoaffiliation
{Frontier Science Center for Quantum Information, Beijing 100084, P.R. China}
\author{Peizhe Tang}
\affiliation
{School of Materials Science and Engineering, Beihang University, Beijing 100191, P.R. China}
\alsoaffiliation
{Max Planck Institute for the Structure and Dynamics of Matter, Center for Free-Electron Laser Science, Hamburg 22761, Germany}
\email{peizhet@buaa.edu.cn}
\author{Wenhui Duan}
\affiliation
{State Key Laboratory of Low Dimensional Quantum Physics and Department of Physics, Tsinghua University, Beijing 100084, P.R. China}
\alsoaffiliation
{Frontier Science Center for Quantum Information, Beijing 100084, P.R. China}
\alsoaffiliation
{Institute for Advanced Study, Tsinghua University, Beijing 100084, P.R. China}
\author{Shuyun Zhou}
\affiliation
{State Key Laboratory of Low Dimensional Quantum Physics and Department of Physics, Tsinghua University, Beijing 100084, P.R. China}
\alsoaffiliation
{Frontier Science Center for Quantum Information, Beijing 100084, P.R. China}
\email{syzhou@mail.tsinghua.edu.cn}

\title
{Occupation Dynamics of Floquet-Volkov States and Spectral Sum Rule}


\begin{document}


\begin{tocentry}
	\centering
	\includegraphics[width=8.2 cm]{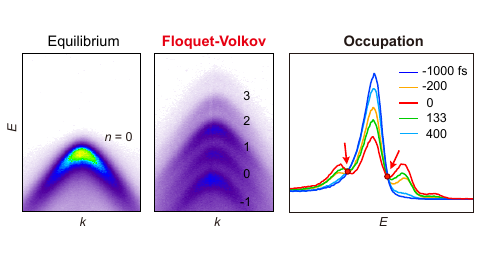}
	
\end{tocentry}


\begin{abstract}
Time-periodic light fields can dress electronic states in quantum materials, forming Floquet states whose dynamic occupation determines transient material properties. Here by using time- and angle-resolved photoemission spectroscopy (TrARPES), we reveal the transient occupation of Floquet-Volkov states in two semiconductors, black phosphorus and MoSe$_2$. While the occupation of the light-induced sidebands, directly reflected by TrARPES spectral weight, strongly depends on the driving field, we find that the total spectral weight obtained by summing up all sidebands is conserved upon below-gap driving. Our work provides critical insights into the Floquet population dynamics, which are essential for light-field tailoring of transient material properties.
\end{abstract}


Floquet engineering, the control of quantum materials through time-periodic drive, has emerged as a powerful paradigm for creating non-equilibrium states of matter with tailored properties \cite{oka2019review,Hsieh2017review,Lindner2020review,Sentef2021review,SYZhou2022review}. This approach is rooted in a fundamental analogy with the Bloch states \cite{AshcroftMermin}: the spatially-periodic potential in crystals leads to Bloch bands which are periodic in the momentum space. Similarly, a time-periodic drive dresses the Bloch states inside the crystal, forming energy-periodic Floquet-Bloch states \cite{shirley1965floquet,Oka2009PhotoHE}. Such light-field dressed states, $|\psi_\alpha(t)\rangle$, are superpositions of multiple ``photon-dressed'' states $|u_{\alpha}^n\rangle$, expressed as $|\psi_\alpha(t)\rangle = \sum_{n} |u_{\alpha}^n\rangle e^{-i(\epsilon_\alpha + n\hbar\omega)t / \hbar}$. The eigen energies of these constituent states are different by integer multiples of the drive photon energy $n\hbar\omega$. The interaction of these Floquet states with the time-periodic potential can further open up energy gaps and modify the non-equilibrium properties of target materials, leading to light-field tailored electronic structure \cite{WangYH2013science,SYZhou2023nature,syzhou2026NM,syzhou2026CPL,mahmood2025floquet,dani2026np,syzhou2025TiSe2} and emergent phenomena such as the light-induced anomalous Hall effect \cite{Eugene2011photoHALL, Cavalleri2020photoHALL}, the optical Stark effect \cite{Gedik2015Stark, Gediksci2018, Ghimire2023exciton}, and tailored optical nonlinearities \cite{Hsieh2021TrSHG, Kogar2024TrSHG}. 

Crucially, the resulting transient electronic, transport, and optical properties are governed by the non-equilibrium occupation of these Floquet states \cite{Eugene2011photoHALL,sato2019microscopic,sato2019light, QNiu2022prr}. Unlike the equilibrium state, where bands below the Fermi energy are completely filled, the Floquet regime creates a distinct electronic structure, where the valence band (VB) and its multiple sidebands, both above and below the Fermi energy, coexist as a coherent superposition. This results in a complex population landscape where each state is only partially filled and dynamically evolving with the drive \cite{Refael2015PRX,seetharam2018absence,Claudio2015prb,Huber2023nature}.
In addition, as solid-state materials under laser driving are open systems coupled with baths, the Floquet occupation dynamics often involves complicated competition between the laser driving and the relaxation of photo-excited electrons in the target materials \cite{sato2019light,sato2019microscopic,Claudio2015prb,kohn2001periodic,matsyshyn2023fermi,Young2013prb,liu2015classification,iadecola2015floquet,hone2009statistical,Refael2015PRX,liu2017keldysh,shirai2015condition,syzhou2025prl}. Thus, in sharp contrast to equilibrium quantum statistics, which guarantees the conservation of total electron spectral weight, the total occupation of these Floquet sidebands in non-equilibrium is not always guaranteed to be conserved \cite{matsyshyn2023fermi,Refael2015PRX} due to strong coupling between these states and the baths [see discussions in Supporting Information (SI)]. 
Therefore, it is critical to characterize the intricate population dynamics and investigate under what conditions a spectral sum rule can be established.

Here we directly reveal the population dynamics of light-field dressed states in two semiconductors, black phosphorus and 2H-MoSe$_2$, by using time- and angle-resolved photoemission spectroscopy (TrARPES). Under intense pumping, we observe sidebands up to the third order, and a complete depletion of the VB. Interestingly, while the occupation of each sideband strongly depends on the driving field and the delay time, we observe isosbestic points in the TrARPES spectra upon below-gap pumping, indicating a spectral sum rule, namely, conservation of the total TrARPES spectral weight at a fixed momentum. Such a sum rule is violated by above-gap pumping that involves direct optical transitions in addition to coherent light-matter coupling. Our work provides fundamental insights into the occupation dynamics of Floquet states, which are crucial for the modulation of ultrafast transport and opto-electronic properties in light-engineered quantum materials.

\begin{figure*}[htbp]
	\includegraphics[width=16 cm]{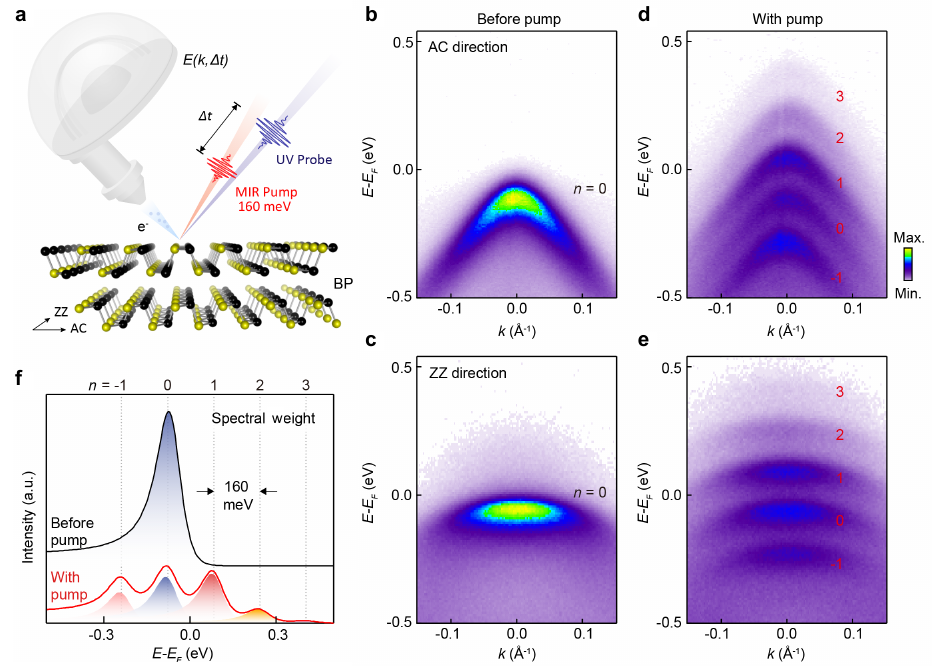}
	\caption{Light-induced sidebands and spectral weight redistribution in black phosphorus upon below-gap pumping.
	(a) Schematic of TrARPES setup with MIR pumping.
	(b,~c) Dispersion images measured at $\Delta t$ = $-$1 ps along the AC (b) and ZZ (c) directions.
	(d,~e) Dispersion images measured at $\Delta t$ = 0  along the AC (d) and ZZ (e) directions.
	(f) EDCs at $k$ = 0 for data in (c,~e). The color-shaded areas are fitting peaks, which are used to extract the spectral weight for each individual peak. The pump has a photon energy of 160 meV and fluence of 0.5 mJ/cm$^2$. The pump is $p$-polarized, with polarization along ZZ direction (ZZ-pump) for (d) and polarization along AC direction (AC-pump) for (e).
    }
    \vspace{0.1cm}
    \label{Fig1}
\end{figure*}

Figure~1a shows a schematic of the TrARPES setup, where a mid-infrared (MIR) pulse at 160 meV is used as the drive (see Figure~S1 in the SI for experimental geometry). 
The equilibrium electronic structures of black phosphorus along the armchair (AC) and zigzag (ZZ) directions are shown in Figure~1b,~c. The dynamic occupation of light-field dressed states is reflected by comparing the TrARPES spectral weight $S(\mathbf{k}) = \int I(\mathbf{k}, E)dE$ with the equilibrium state, where $I(\mathbf{k},E)$ is the TrARPES intensity, $\mathbf{k}$ and $E$ are electron momentum and energy, respectively. The TrARPES intensity is expressed as $I(\mathbf{k},E)\simeq A(\mathbf{k},E) |M(\mathbf{k},E)|^{2} f(\mathbf{k},E)$, where $A$, $f$ denote spectral function and occupation function, respectively, and $M$ is the dipole transition matrix element which is related to measurement geometry and irrelevant to occupation\cite{Damascelli2024RMP}. Under certain geometries, the dipole transition matrix element is similar between different dressed states, so that the extracted spectral weight reveals the dynamic occupation. 

Upon below-gap pumping at a fluence of 0.5 mJ/cm$^2$ (with corresponding peak field strength of 1.1$\times$10$^8$ V/m), the intensity of the VB ($n$ = 0) clearly decreases, and multiple sidebands are clearly observed in Figure~1d,~e. Here, the application of $p$-$pol.$ pump (with non-zero out-of-plane light-field) leads to the interference between the Floquet and Volkov states \cite{mahmood2016NP,bao2025FV,Stefan2024graphene,Gedik2024graphene,fragkos2025floquet}, thereby enhancing sideband intensity with occupation up to the third order ($n$ = 3) (see more details in Figure~S1). In this work, we utilize various combinations of pump and probe polarizations, so that the Floquet optical selection rule guarantees that Floquet states and the related Floquet-Volkov states are clearly observed \cite{SYZhou2024spot,FanBSSciAdv2025}, whose occupations are revealed by TrARPES intensity.

Figure~1f compares the energy distribution curves (EDCs) at the  $\Gamma$ point before and upon pumping, extracted from Figure~1c,~e. The single peak in the equilibrium state becomes weaker upon pumping, with its spectral weight redistributed into sidebands displaced by the drive photon energy of 160 meV. The reduced spectral intensity of the $n$ = 0 VB indicates that there is less electron filling, namely, the VB is only partially occupied, which is in sharp contrast to the occupation of the equilibrium state. 
The spectral weight of these Floquet-Volkov states strongly depends on the pump strength, scaling with the square of the $n$-th Bessel function $J^2_n(\gamma)$. Here, the parameter $\gamma$ depends on the electronic structures of the VBs, the properties of the pumping laser, and the velocity of photoelectrons (see details in Methods). At a pump fluence of 0.5 mJ/cm$^2$, the population is slightly inverted, which is indicated by the slightly higher intensity for the first-order $n$ = 1 sideband (red shaded area in Figure~1f) than the $n$ = 0 band (blue shaded area). Under stronger pumping at 1.1 mJ/cm$^2$, the $n$ = 0 VB is nearly depleted, and the spectral weight is almost entirely redistributed into the $|n|$ $\geq$ 1 sidebands (Figure~S2). Such a drastic population modulation could reshape electronic properties, e.g., by flipping fermionic interactions \cite{tsuji2011PRL}.
In previous works \cite{SYZhou2023nature,SYZhou2023PRL,SYZhou2024glide}, we have demonstrated light-field induced band renormalization, and here we focus on another crucial and fundamental question, the dynamical occupation of these transient sidebands.

The temporal evolution of Floquet-Volkov spectral weight encodes critical information about the underlying transient occupation. Figure~2a-e shows snapshots of dispersion images measured at five representative delay times (marked in Figure~2f). Figure~2g shows the corresponding EDCs at the $\Gamma$ point. Interestingly, the EDCs at different delay times always intersect at two points as pointed by red arrows, which is analogous to isosbestic points in chemical reactions \cite{IP1,IP2,IP3}, Mott transition \cite{Mott1}, and structural transition \cite{gedik2007science} in solid states. Such isosbestic points are indications of the conservation of the total populations where there is inter-conversion between different species/states \cite{IP2,IP3}. The observation of isosbestic points here indicates the inter-conversion between the VB and sidebands with $|n|$ $\geq$ 1, and possible spectral weight conservation (see more discussion in SI). 

\begin{figure*}[htbp]
	\includegraphics[width=16.8 cm]{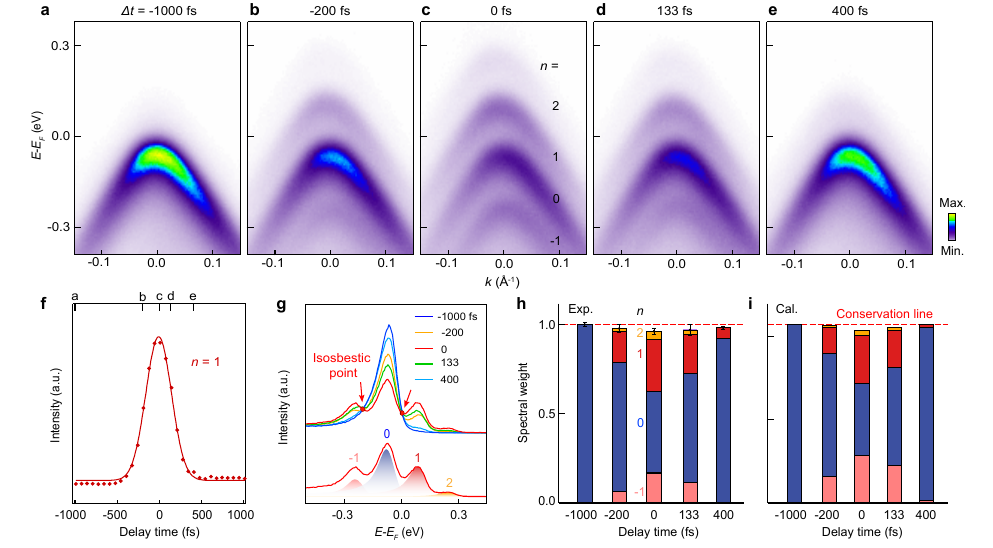}
	\caption{Observation of spectral weight conservation in black phosphorus.
     (a-e) Dispersion images measured along the direction at 30$^{\circ}$ from the AC direction of black phosphorus with $p$-$pol.$ at different delay times. The pump photon energy is 160 meV, and the pump fluence is 0.32 mJ/cm$^2$.
	(f) Intensity of $n$ = 1 sideband as a function of the delay time.
	(g) EDCs at $k$ = 0 from data in (a-e), and fitting of the EDC at $\Delta t$ = 0 (bottom curve).
	(h) Extracted spectral weight for different sidebands at different delay times from EDCs in (g). The error bars are for the total spectral weight and are estimated via standard uncertainty propagation from the fitting parameters.
	(i) Calculated spectral weight for $n = 0, \pm1, 2$ sidebands at different delay times.
    }
	\label{Fig2}
\end{figure*}

The Floquet-Volkov spectral weight $S_{FV}(\textbf{k})$ is experimentally defined as the total TrARPES spectral weight by summing up the spectral weight of the VB $S_{n=0}(\textbf{k})$ and its sidebands $S_{n\neq 0}(\textbf{k})$ at a fixed momentum \textbf{k}, 
\begin{equation}
S_{FV}(\textbf{k})=\sum_{n}S_{n}(\textbf{k})
\end{equation} 
The spectral weight for each sideband $S_n(\textbf{k})$ is extracted by integrating TrARPES intensity $I_n(\textbf{k},E)$ over electron energy $E$ (shaded area in Figure~2g). Figure~2h shows the extracted spectral weight for each sideband at different delay times, which changes with delay time due to the different field strength. Interestingly, the total spectral weight, summed over all sidebands ($n$ = 0 and $n$ $\neq$ 0), is nearly conserved and matches the unpumped spectral weight (red dashed line). Theoretically, we use the Floquet theory to describe Floquet-Volkov interference, where the spectral weight of each sideband is simplified as a Bessel function without considering the coupling with the baths (see details in SI). Under this approximation, we model the redistribution of spectral weight among different sidebands and reproduce its conservation, as shown in Figure~2i. 
The qualitative agreement between simulated results and TrARPES measurements shows that a spectral sum rule holds, reflecting conserved occupations of light-dressed states. The conservation of total spectral weight and the presence of isosbestic points are also observed for different pump fluences (Figure~S3) and in pure Floquet states under $s$-$pol.$ pumping with weaker sidebands (Figure~S4). Moreover, the spectral sum rule also holds for non-high-symmetry $k$ points in addition to the $\Gamma$ point (Figure~S5).
 
The spectral sum rule is also observed in 2H-MoSe$_{2}$ upon pumping at 311 meV, which is far below its band gap of 1.1 eV \cite{michael2012tmd}. Figure~3a shows an overview of the electronic structure using an advanced high-photon-energy light source \cite{ultrafastsci}. Clear sidebands are observed around the K valley for both $s$-$pol.$ and $p$-$pol.$ pump (Figure~3b-g). In both cases, the pump polarization has a non-zero out-of-plane field component, which is larger for $p$-$pol.$ pump, leading to stronger Floquet-Volkov sidebands in Figure~3d. 
Interestingly, a similar EDC analysis at the K point shows that the total spectral weight is conserved (Figure~3h,~i), suggesting that the spectral sum rule also holds for 2H-MoSe$_2$ upon below-gap pumping.

\begin{figure*}[htbp]
	\includegraphics[width=1\textwidth]{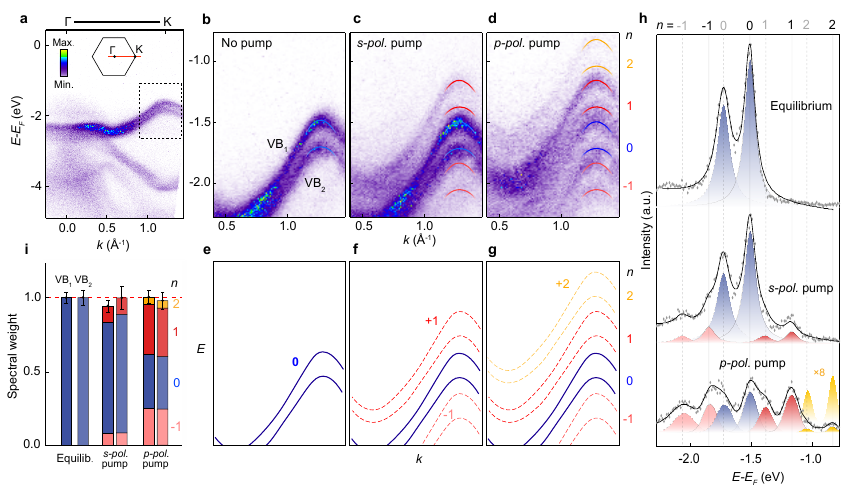}
  \caption{Observation of spectral weight conservation in MoSe$_{2}$.
    (a) Dispersion image of bulk MoSe$_{2}$ measured along the $\Gamma$-K direction. The dashed box marks the band structure at the K valley, which is the main focus.
	(b-d) Dispersion images measured around the K valley without pump (b), with $s$-$pol.$ (c) and $p$-$pol.$ (d) pump. The pump photon energy is 311 meV and the pump fluence is 2.5 mJ/cm$^2$.
	(e-g) Corresponding dispersions for data shown in (b-d).
	(h) EDCs at the K point for data shown in (b-d) and fitting results to extract the individual spectral weight.
	(i) Extracted spectral weight for different sidebands from EDC analysis in (h). 
    }
    \label{Fig3}
\end{figure*}

We further explore the spectral sum rule by tuning pump photon energy across the band gap. Our results reveal a breakdown of the spectral sum rule for the light-induced valence sidebands once an optical transition to the conduction band (CB) is activated. Figure~4 shows a comparison of the spectral weight for valence sidebands of black phosphorus under below-gap (see schematic in Figure~4a) to above-gap pumping (Figure~4b). For below-gap pumping, the dispersion images at various pump photon energies show that the Floquet-Volkov spectral weight is conserved (Figure~4c-g). In contrast, upon above-gap AC-pumping (Figure~4h-k), such conservation is violated, which is supported by the analysis in Figure~4p (indicated by blue arrows). This breakdown occurs because above-gap AC-pump activates photo-excitation into the CB in accordance with the optical selection rule \cite{Kim2020NatMat}. As shown by the population of CB in Figure~4l-o and Figure~S6, electrons are not only redistributed among the sidebands, but are also transiently excited to the CB. This additional excitation channel is responsible for the non-conservation of the total Floquet-Volkov spectral weight for the valence sidebands. In addition, we also note that for above-gap ZZ-pump where the direct interband transition is forbidden\cite{Kim2020NatMat}, the Floquet-Volkov spectral weight is conserved (Figure~S7).

\begin{figure*}[htbp]
	\includegraphics{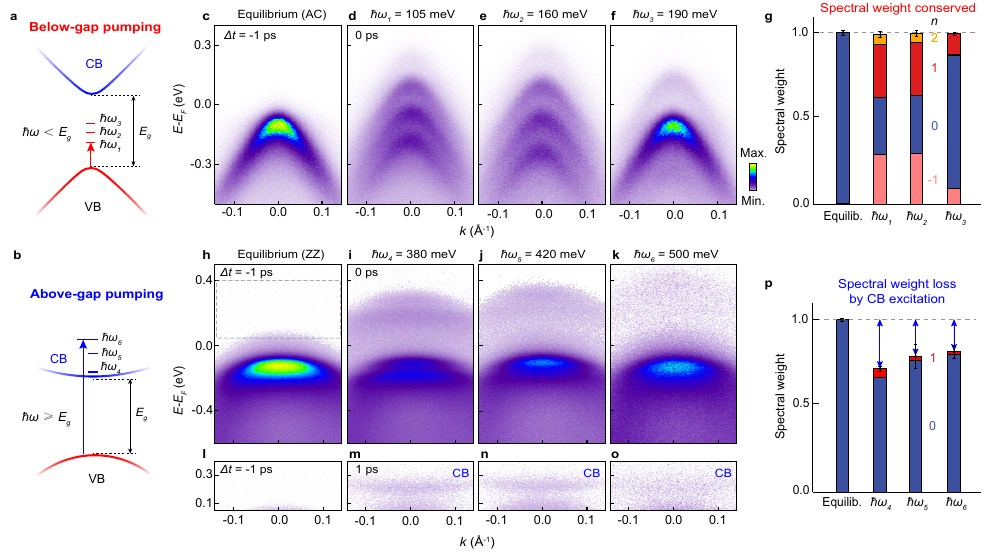}
	\caption{Conservation of spectral sum rule upon below-gap pumping, and violation of spectral sum rule upon above-gap pumping in black phosphorus.
	(a,~b) A schematic illustration for the band structure and pump photon energies below the band gap $E_{g}$ (a) and above the band gap (b).
	(c-f) Dispersion images measured along AC direction at $\Delta t$ = $-$1 ps (c) and at $\Delta t$ = 0 ps with three different pump photon energies (d-f). The pump is ZZ-pump ($p$-$pol.$) and the pump fluences are 0.29, 0.26 and 0.20 mJ/cm$^{2}$, respectively.
	(g) Extracted spectral weight at the $\Gamma$ point from data in (c-f).
	(h-k) Dispersion images measured along the ZZ direction at $\Delta t$ = $-$1 ps (h) and at $\Delta t$ = 0 ps with three different pump photon energies larger than the band gap (i-k). The pump is AC-pump ($p$-$pol.$) and the pump fluence is 0.7 mJ/cm$^{2}$.
	(l-o) Zoomed-in dispersion images at $\Delta t$ = $-$1 ps (l) and at $\Delta t$ = 1 ps for three different above-gap pump photon energies (m-o).
	(p) Extracted spectral weight at the $\Gamma$ point from data in (h-k).}\label{Fig4}
\end{figure*}

The dynamic occupation of Floquet-Volkov states is manifested in the transfer of spectral weight among Floquet-Volkov sidebands and the conservation of the total spectral weight. Within the Floquet-Volkov description, the $n$-th sideband is characterized by the Floquet-Volkov coefficient $c_n$, with a sideband weight given by $|c_n|^2=J_n^2(|\gamma|)$. Here, $\gamma=\beta-\alpha$ is a dimensionless parameter that characterizes the interference between the Floquet and Volkov states, where $\beta$ describes the Floquet dressing of the initial state and $\alpha$ the Volkov dressing of the photoelectron (see details in Methods). The $n$-th Bessel function $J_n$, which are tunable by the pumping strength, corresponds to the amplitudes of the harmonics of the Floquet wave functions \cite{matsyshyn2023fermi}. Under below-gap pumping, where single photon absorption is suppressed, the sum of $J^2_n$ over the Floquet index $n$ is unity, as observed in both black phosphorus and MoSe$_{2}$. In contrast, above-gap pumping may activate direct optical transitions, populating the CB with carriers and thereby diminishing the occupation of the VB and its sidebands. This leads to the violation of the spectral sum rule. We note an exception in black phosphorus when measured along the AC direction with AC-pump:  even for below-gap pumping, the total spectral weight is not conserved around the $\Gamma$ point (Figure~S8). This arises from the Floquet matrix element effect \cite{SYZhou2024spot,FanBSSciAdv2025}, which modulates the TrARPES intensity such that the measured spectral weight deviates from the true electronic occupation. This is purely a detection-related phenomenon; in principle, the total occupation of the Floquet–Volkov states still remains conserved. We also note that in the sub-cycle regime, the coherent light field can induce momentum shifts of spectral weight or the entire bands\cite{Huber2018nature,Huber2023nature,Ofer2022PRR,Rubio2024JPCM}, potentially leading to a breakdown of the sum rule. Such effects require ultrashort probe pulses and are beyond the scope of the present study.

In summary, we reveal the occupation dynamics of Floquet-Volkov states in two semiconductors. We find that below-gap pumping conserves the total spectral weight, establishing a spectral sum rule. This conservation indicates that electron occupation is preserved during light-field dressing, involving interconversion between the $n$ = 0 and  $n$ $\neq$ 0 states, and suggests negligible bath-mediated particle exchange out of the valence band Floquet manifold on the ultrafast timescale under below-gap pumping, although coherent energy exchange with the driving field remains essential. In contrast, above-gap pumping may open an alternative channel by directly exciting electrons into the CB, which diminishes the population of the light-dressed sidebands and breaks the spectral sum rule. 
Spectral sum rules are important across different fields of physics, from particle physics \cite{particlesum1} to condensed matter physics \cite{conductsum, XMCDsum1}. For instance, the sum rule in X-ray magnetic circular dichroism (XMCD) is used to determine the orbital or spin moments \cite{XMCDsum1}. Our demonstration of the Floquet spectral sum rule, combined with the direct observation of transient occupation of light-field dressed states, suggests that below-gap pumping creates pure light-field dressed states, and provides useful insights for understanding and controlling the ultrafast transport and optoelectronic properties of Floquet quantum materials.


\section{METHODS}
\subsection{Sample preparation}
High-quality single crystals of black phosphorus and MoSe$_{2}$ were grown by chemical vapour transport method.  For black phosphorus, a mixture of red phosphorus lump (Alfa Aesar, 99.999$\%$), tin grains (Aladdin, $\geq$ 99.5$\%$), and iodine crystals (Alfa Aesar, 99.9$\%$) was heated to 600$^{\circ}$C for 1 day in an evacuated silica tube. For MoSe$_{2}$, Mo foil (Alfa Aesar, 99.95$\%$) and Se ingot (Alfa Aesar, 99.99$\%$) were mixed and heated to 900$^{\circ}$C for 3 days. The as-grown MoSe$_{2}$ was then recrystallized using SeCl$_{4}$ (Aladdin, 99$\%$) as the transporting agent under 980$^{\circ}$C for 9 days. Millimeter-sized black phosphorus and MoSe$_{2}$ single crystals were obtained.

\subsection{TrARPES measurements}
TrARPES measurements were performed in the home laboratory at Tsinghua University with a regenerative amplifier laser with a center wavelength of 800 nm (1.55 eV) and a pulse energy of 1.3 mJ, at a repetition rate of 10 kHz. The majority of the beam is used to drive the optical parametric amplifier. The MIR pump beam is generated by non-collinear differential frequency generation of the signal and idler of the optical parametric amplifier. The probe beam with a photon energy of 6.2 eV is generated by a three-step fourth-harmonic generation process using beta barium borate crystals. The temporal scales of the MIR pump and the 6.2 eV probe are shown in Figure~S9. The probe beam with a photon energy of 21.7 eV is generated via high harmonic generation by focusing the second harmonic (3.1 eV) of the fundamental beam into an Argon-filled gas cell \cite{ultrafastsci}. The 21.7 eV beam (7th harmonic of 3.1 eV) is isolated from other harmonics by passing through an aluminium foil and a tin foil.
The samples were cleaved and measured at a temperature of 80 K in an ultra-high vacuum chamber with a base pressure better than 5 $\times$ $10^{-11}$ Torr. 

\subsection{The extraction of spectral weights}
The spectral weights are extracted by fitting the EDCs with Lorentzian peaks multiplied by a Fermi–Dirac function plus a linear background. The EDCs are obtained by integrating over a momentum window of 0.02 $\si{\angstrom^{-1}}$ around each $k$ point. The spectral weight of each sideband $S_{n}(\mathbf{k})$ is determined by integrating the corresponding fitting peak $I_{n}(\mathbf{k}, E)$ over energy $E$. The associated error bars are estimated via standard uncertainty propagation from the fitting parameters.

\subsection{The simulation of spectral weights}
In this work, we employ Floquet theory to calculate the spectral weight of the Floquet-Volkov state in black phosphorus. We begin by deriving the spectral weight for the Floquet and Volkov states, which subsequently allows us to determine the spectral weight of the Floquet-Volkov state \cite{Park2014PRA}.

For the Floquet state, we consider the in-plane component of the vector potential $\mathbf{A}_{\|}(t)$ of the pumping laser as
\begin{equation}
\label{pump}
	\begin{aligned}
		\mathbf{A}_{\|}(t)&=A_x(t)\hat{\textbf{x}}+A_y(t)\hat{\textbf{y}}\\
		&=A_x\cos(\omega t)\hat{\textbf{x}}+A_y\cos(\omega t)\hat{\textbf{y}}
	\end{aligned}
\end{equation}
where $A_x$ and $A_y$ are the amplitudes of the vector potential along $x$ (ZZ) and $y$ (AC) directions, and $\omega$ is the frequency of the pumping laser. We consider the electronic structures of black phosphorus in equilibrium to be well described by a parabolic band and incorporate the influence of the pumping laser using the Peierls substitution. Therefore, the effective Hamiltonian $\hat{H}_{\Gamma}(t,p_{\|})$ of the VB edge around the $\Gamma$ point is given by
\begin{equation}
\begin{aligned}
\hat{H}_{\Gamma}(t,p_{\|}) &=\sum_{i=x, y} \frac{\left[p_{i}+e A_{i}(t)\right]^{2}}{2 m_{i}}\\&=\sum_{i=x, y}\left[\frac{p_{i}^{2}}{2 m_{i}}+\frac{e p_{i} A_{i}(t)+e A_{i}(t) p_{i}}{2 m_{i}}+\frac{e^{2} A_{i}^{2}(t)}{2 m_{i}}\right] \\
&\simeq\sum_{i=x, y}\left[\frac{p_{i}^{2}}{2 m_{i}}+\frac{e A_{i}(t) p_{i}}{m_{i}}\right]
\end{aligned}
\end{equation}
where $e$ is the charge of electron, $p_{i}=\hbar k_i$ is the lattice momentum with $\hbar$ as the reduced Planck constant, and $m_{i}$ is the effective mass of the electron along the $i$ direction. The higher-order terms are neglected. Applying eq~\ref{pump}, we can simplify $\hat{H}_{\Gamma}(t,p_{\|})$ as

\begin{equation}
\begin{aligned}
\hat{H}_{\Gamma}(t,p_{\|})&=\sum_{i=x, y} \frac{p_{i}^{2}}{2 m_{i}}+\frac{e A_{x}(t)p_{x}m_y+eA_{y}(t)p_ym_x}{m_{x}m_y}\\
&=\sum_{i=x, y} \frac{p_{i}^{2}}{2 m_{i}}+\hbar\omega \beta \cos (\omega t)\\
&=\hat{H}_{0}(p_{\|})+\hat{H}_{I}(t, \beta)
\end{aligned}
\end{equation}
where $\beta=\frac{e (A_{x}p_xm_y+A_{y}p_ym_x)}{\hbar\omega m_{x}m_y}$. Using the Jacobi-Anger relation $e^{-in\sin\theta}=\sum_mJ_m(n)e^{-im\theta}$, we find that the evolution of the Floquet state is governed by $\hat{H}_I(t,\beta)=\hbar\omega \beta\cos (\omega t)$, leading to a wavefunction of the Floquet state given by
\begin{equation}
\begin{aligned}
\Psi^{F}(t)&=\exp \left[-\frac{i}{\hbar} \int d t \hat{H}_{I}(t,\beta)\right] \Psi(t_0)\\
&=\exp \left[-i\beta \sin (\omega t)\right] \Psi(t_0) \\
&
=\Psi(t_0) \sum_{m=-\infty}^{+\infty} J_{m}\left(\beta\right) e^{-i m \omega t}
\end{aligned}
\end{equation}
where $\Psi(t_0)$ is the unperturbed wavefunction and $J_{m}(x)$ is the $m$-th Bessel function of the first kind. Now we define the Floquet coefficient $b_{m}$ as $J_{m}(\beta)$. If we do not consider the dissipative effects \cite{mahmood2016NP}, the spectral weight of the $m$-th Floquet sideband is $I_m^F=|b_m|^2$.

The Volkov state arises from the interference between the pumping laser and photoexcited electrons, leading to a Hamiltonian of $\hat{H}_{V}(t)=e v_{z} A_z(t)=e v_{z} A_z\cos(\omega t)$, where $v_z$ is the out-of-plane electron velocity and $A_z$ is the vertical component of the vector potential of the pumping laser. Similar to the Floquet state, we obtain the time-dependent evolution of the Volkov state as
\begin{equation}
\begin{aligned}
\Psi^{v}(t)&=\exp \left[-\frac{i}{\hbar} \int d t\hat{H}_V(t)\right] \Psi_{k}(t_0)\\
&=\exp \left[-\frac{i}{\hbar} \frac{e A_{z}v_{z} \sin (\omega t) }{\omega}\right] \Psi_{k}(t_0)\\
&=\Psi_{k}(t_0) \sum_{n=-\infty}^{+\infty} J_{n}(\alpha) e^{-i n \omega t}
\end{aligned}
\end{equation}
where $\Psi_{k}(t_0)$ is the free electron wavefunction and $\alpha=\frac{e A_{z} v_{z}}{\hbar \omega}$. The Volkov coefficient $a_{n}$ is defined as $J_{n}(\alpha)$. For the Volkov state, the spectral weight of the $n$-th Volkov energy level is $I_n^V=|a_n|^2$.

For the Floquet-Volkov state, the interference between Floquet states and Volkov states should be considered. Herein, we can define the Floquet-Volkov coefficient \cite{Park2014PRA} as
\begin{align}
c_{q} \equiv \sum_{n, m}^{m-n=q} a_{n}^{\dagger} b_{m}=\sum_{n^{\prime}, m}^{m+n^{\prime}=q} a_{-n^{\prime}}^{\dagger} b_{m}
\end{align}
where the $(n, m)$ summation is a discrete convolution and $a_{-n^{\prime}}^{\dagger}=(-1)^{n^{\prime}} a_{n^{\prime}}$. Using the summation theorem for the Bessel function of the first kind \cite{gradshteyn2014table}, the Floquet-Volkov coefficient becomes
\begin{equation}
\begin{aligned}
c_{q} &=\sum_{n^{\prime}, m}^{m+n^{\prime}=q}(-1)^{n^{\prime}} J_{n^{\prime}}(\alpha)  J_{m}\left(\beta\right)\\
&=\sum_{n^{\prime}} J_{-n^{\prime}}(\alpha) J_{q-n^{\prime}}\left(\beta\right)\\
&=\sum_{n^{\prime}} J_{n^{\prime}}(\alpha) J_{q+n^{\prime}}\left(\beta\right)\\
&=\left(\frac{\gamma}{|\gamma|}\right)^{q} J_{q}(|\gamma|)
\end{aligned}\label{cq}
\end{equation}
where $\gamma =\beta-\alpha$.
So for the Floquet-Volkov state, the spectral weight of the $q$-th sideband is $I_q^{FV}=|c_q|^2$.

In our calculations of the spectral weight for the Floquet-Volkov state (see Figure~2i in the main text and Figure~S3h), the photon energy of the pumping laser is set at $\hbar\omega=160$ meV, and the effective mass of the electron $m_x$ $(m_y)$ along the $x$ ($y$) direction is 0.81$m_e$ (0.13$m_e$), where $m_e$ is the electron mass. For pump-probe experiments with different delay times upon ZZ-pump ($p$-$pol.$, see Figure~S1a for experimental geometry), we employ the electric field as $E(t)=(E_x,E_z)e^{-\frac{(t/\tau)^2}{2}}$, where $\tau$ = 222 fs and $t$ represents different delay times as $-$1000, $-$200, 0, 133 and 400 fs. The parallel and vertical electric field amplitudes $E_x$ and $E_z$ with respect to the sample surface are  $3.35\times10^7$ V/m and $6.4\times10^7$ V/m, respectively. The vector potential of the pumping laser $A_{x,z}$ is calculated via $E_{x,z} / \omega$, and the electron velocity $v_z$ in the out-of-plane direction is $7.7\times10^5$ m/s. In this case, the corresponding $|\gamma|$ is 0, 0.844, 1.267, 1.059, and 0.250 at different delay times.

To simulate Floquet-Volkov spectral weight under different pump fluences (as shown in Figure~S3h), we determine the vector potential $A_{x,z}$ from the pump fluences $F$ used in the pump-probe experiments ($F$ = 0, 0.081, 0.156, 0.207 and 0.276 mJ/cm$^2$), with the corresponding $|\gamma|$ as 0, 0.640, 0.889, 1.024 and 1.182, respectively.



\begin{suppinfo}
	This material is available free of charge via the internet at http://pubs.acs.org. \\
	Experimental geometry, complete occupation depletion of the original valence band, Floquet spectral sum rule holds under different pump fluences, Floquet spectral sum rule for pure Floquet-Bloch states, underlying mechanism for observing isosbestic points in the Floquet spectrum, conservation of spectral weight for both high-symmetry and non-high-symmetry $k$ points, spectral weight dynamics upon above-gap AC-pumping in black phosphorus, conservation of spectral weight for above-gap ZZ-pumping in black phosphorus, violation of Floquet spectral sum rule due to Floquet matrix element effect, temporal scales of the pump and probe pulses, and more discussions on the spectral weights. (PDF).
\end{suppinfo}

\section{AUTHOR INFORMATION}
\subsection{Author Contributions}
$^{\infty}$ X.C., C.B. and B.F. contributed equally to this work.

Shuyun Z. conceived the research project. X.C., C.B., Haoyuan Z. and Shaohua Z. performed the TrARPES measurements and analyzed the data. F.W. and Haoyuan Z. grew the black phosphorus and MoSe$_{2}$ single crystals. B.F., P.T. and W.D. performed the theoretical analysis and calculation. T.L., Hongyun Z. and P.Y. contributed to the data analysis and discussions. X.C. and Shuyun Z. wrote the manuscript, and all authors contributed to the discussions and commented on the manuscript.

\subsection{Notes}
The authors declare no competing financial interest.

\subsection{Acknowledgment}
This work is supported by the National Natural Science Foundation of China (Grants No.~12234011, 12421004), Tsinghua University
Initiative Scientific Research Program (Grant No.~20251080106), National Natural Science Foundation of China (52388201, 12327805, 12374053), the National Key Basic Research and Development Program of China (Grants No. 2024YFA1409100), and the New Cornerstone Science Foundation through the XPLORER PRIZE.


\bibliography{reference}


\end{document}